\documentclass[namedreferences,hyperref,optionalrh,solaromanenum]{spr-sola}

\usepackage{graphicx}                    
\usepackage{amssymb}                    
\usepackage{amsmath}
\usepackage{color}                       
\usepackage{natbib}
\usepackage{lineno}

\newcommand{\etal}{{\it et al.}}
\newcommand{\eg}{{\it e.g.}}
\newcommand{\ie}{{\it i.e.}}
\newcommand{\sdo}{{\it SDO}} %
\newcommand{\hmi}{\sdo/HMI} %
\newcommand{\degr}{\mbox{$^\circ$}}%
\newcommand{\dt}{\mbox{$\delta\tau$}} %
\newcommand{\dtdl}{\mbox{$\delta\tau(\Delta, \lambda)$}}
\newcommand{\dthw}{\mbox{$\delta\tau_\mathrm{H}^\mathrm{W}$}} %
\newcommand{\dtgw}{\mbox{$\delta\tau_\mathrm{G}^\mathrm{W}$}} %
\newcommand{\dthe}{\mbox{$\delta\tau_\mathrm{H}^\mathrm{E}$}} %
\newcommand{\dtge}{\mbox{$\delta\tau_\mathrm{G}^\mathrm{E}$}} %

\begin{document}
\hypersetup{
    colorlinks=true,
    linkcolor=blue,    
    citecolor=blue,    
    urlcolor=blue      
}
\begin{frontmatter}


\title{Oscillatory Phase and Acoustic Travel-Time Inconsistencies Measured between SDO/HMI and GONG Dopplergrams}
\author[addressref=aff1,corref,email={junwei@sun.stanford.edu}]{\inits{J.}\fnm{Junwei}~\lnm{Zhao}}\orcid{orcid:0000-0002-6308-872X}
\author[addressref=aff1]{\inits{R.}\fnm{Ruizhu}~\lnm{Chen}}\orcid{orcid:0000-0002-2632-130X}
\author[addressref=aff2]{\fnm{S.~P.~}\lnm{Rajaguru}}\orcid{orcid:0000-0003-0003-4561}
\author[addressref={aff3,aff4}]{\inits{S.}\fnm{Shukur}~\lnm{Kholikov}}\orcid{orcid:0000-0003-1860-3697}

\address[id=aff1]{W.~W.~Hansen Experimental Physics Laboratory, Stanford University, Stanford, CA 94305-4085, USA}
\address[id=aff2]{Indian Institute of Astrophysics, II Block Koramangala, Bengaluru 560 034, India}
\address[id=aff3]{National Solar Observatory, 3665 Discovery Dr., Boulder, CO, 80303, USA}
\address[id=aff4]{National Research University TIIAME, Kori Niyoziy 39, Tashkent 100000, Uzbekistan}

\runningauthor{Zhao \etal}
\runningtitle{GONG and HMI Phase Anomalies}

\begin{abstract}
We investigate the causes of discrepancies in meridional-circulation measurements derived from the helioseismic observations by the Helioseismic and Magnetic Imager onboard the {\it Solar Dynamics Observatory} (\hmi) and the Global Oscillation Network Group (GONG). 
Using contemporaneous Dopplergrams from both instruments that are processed consistently, we measure relative oscillatory phase shifts at identical solar locations and analyze north-south acoustic travel-time shifts on both sides of the solar central meridian. 
Our analysis reveals a persistent area of phase-shift anomalies in the northwestern quadrant of the solar disk, whose magnitude increases over the analysis period from 2010 through 2024. 
After removing the axisymmetric component, the phase-shift maps display a deceasing trend from the northeastern to the southwestern quadrant, which can be misinterpreted as flows in helioseismic analyses. 
The travel-time measurements also show significant inconsistencies on the eastern and western sides of the central meridian for both instruments, although a close agreement between the both sides is expected. 
These findings indicate that both \hmi\ and GONG carry systematic artifacts affecting meridional-circulation measurements, and that the time-varying phase anomalies and eastern-western asymmetry pose major challenges for their accurate characterization and correction.
\end{abstract}


\end{frontmatter}

\section{Introduction}
\label{sec1}

Meridional circulation at the solar surface and within the solar interior plays a crucial role in the solar dynamo, transporting magnetic flux and redistributing angular momentum across latitudes \citep[\eg,][]{Charbonneau2020, Choudhuri2021}. 
Although the near-surface meridional flow has been extensively measured \citep[\eg,][]{Haber2002, Zhao2004, Hathaway2010, Gonzalez2010, Komm2015}, the internal circulation profile remains a subject of ongoing debate. 
Due to the relatively slow speed of meridional flow inside the convection zone and the significant systematic center-to-limb (CtoL) effect associated with the helioseismic measurements \citep[\eg,][]{Zhao2012, Baldner2012, Chen2018}, inferring the Sun’s internal meridional circulation is particularly challenging. 
Consequently, studies by different authors have yielded inconsistent results.

Using Dopplergrams obtained by the Helioseismic and Magnetic Imager \citep[HMI;][]{Scherrer2012, Schou2012} aboard the Solar Dynamics Observatory \citep[\sdo;][]{Pesnell2012}, the Michelson Doppler Imager \citep[MDI;][]{Scherrer1995} aboard the Solar and Heliospheric Observatory \citep[{\it SOHO};][]{Domingo1995}, and the Global Oscillation Network Group \citep[GONG;][]{Harvey1996}, several research teams have applied time--distance helioseismology techniques \citep{Duvall1993} to measure the internal meridional circulation. 
While the measurements and inversions by \citet{Zhao2013}, \citet{Chen2017}, and \citet{Lin2018} support the double-cell circulation pattern in each solar hemisphere, the findings of \citet{Jackiewicz2015}, \citet{Rajaguru2015}, and \citet{Gizon2020} favor a single-cell configuration. 
Other researchers have suggested that current observational techniques and measurement precision are insufficient to definitively distinguish between single-cell and double-cell circulation patterns \citep{Boening2017, Stejko2022, Hatta2024}.

While the helioseismic CtoL effect is a significant factor contributing to the discrepancies among different authors’ measurements, \citet{Gizon2020} pointed out that a systematic instrumental effect may also be present in one of the observing instruments. 
By analyzing Doppler observations from all three aforementioned instruments, they found that the meridional flow derived from \hmi\ Dopplergrams consistently exhibited a stronger southward flow compared to those from the other two instruments. 
This discrepancy led them to conclude that the \hmi\ instrument might be subject to a systematic bias. 
Subsequent Fourier-Legendre decomposition analysis \citep{Braun1998, Roth2016} confirmed this finding \citep{Braun2021}.
\citet{Chen2025} further noted that while meridional flow measurements using \hmi\ and GONG data produce generally consistent results, a persistent offset exists in the measured acoustic travel times. However, it remains unclear which instrument is responsible for this systematic offset despite previous claims.

To reconcile the discrepancies in meridional-circulation profiles reported by different research teams using data from different instruments, it is essential to investigate the origin(s) of the travel-time offset found in the measurements using \hmi\ and GONG data. 
A better understanding of this anomaly may allow for appropriate and more accurate corrections in future analyses.
In this paper, we analyze contemporaneous Dopplergrams from \hmi\ and GONG to investigate this systematic discrepancy through comparing the disk-location-dependent oscillatory phases and the measured acoustic travel times. 
Section~\ref{sec2} describes the datasets used in this study. 
Section~\ref{sec3} compares the oscillatory phases in the Doppler signals across the solar disk as observed by the two instruments and shows the temporal evolution of the relative phase shifts.
In Section~\ref{sec4}, we present acoustic travel-time measurements on both sides of the central meridian, as well as their temporal evolution.  
We then discuss our findings In Section~\ref{sec5} and present our conclusion in Section~\ref{sec6}.

\section{Data Preparation}
\label{sec2}

The data used in this analysis are prepared following the same procedures as those employed for deep meridional circulation measurements in previous studies \citep{Zhao2013, Rajaguru2015, Chen2017}, and are the same datasets utilized by \citet{Chen2025} in a separate publication.

\begin{figure}
    \centering
    \includegraphics[width=1.0\linewidth]{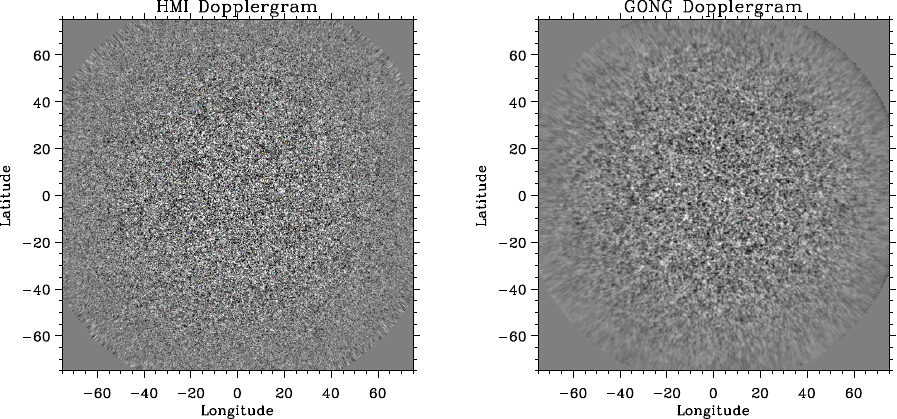}
    \caption{Sample Dopplergrams from \hmi\ ({\it left}) and GONG ({\it right}) after the data are remapped to Postel's projection. }
    \label{dop_images}
\end{figure}

For each day of 14 years from 2010 May 1 to 2024 April 30, 24 hours of continuous full-disk \hmi\ Dopplergrams, acquired at a 45-sec cadence, are remapped to Postel’s projection with a spatial sampling of $0.18\degr$ pixel$^{-1}$. 
The remapped image center is selected as the disk center corresponding to the noon of each day. 
The data were tracked using Snodgrass rate at the disk center \citep{Snodgrass1984}.
Following tracking and remapping, the data were spatially binned down to a resolution of $0.36\degr$ pixel$^{-1}$, resulting in a data cube with dimensions of $425 \times 425 \times 1920$, covering an area of $153\degr \times 153\degr$ on the solar disk. 
GONG Dopplergrams are processed following a similar procedure. 
Although GONG data have lower spatial resolution than \hmi, high spatial resolution is not required for the present analysis, and both sets of data are down-sampled. 
It is also noted that GONG Dopplergrams have a temporal cadence of 60 seconds, which we retain without interpolating them to match the \hmi\ cadence.
Figure~\ref{dop_images} shows two sample remapped Dopplergrams obtained by the \hmi\ and GONG.

Accurate knowledge of the solar $B_0$-angle and $P$-angle is essential for the data preparation. 
Our experience indicates that even small inaccuracies in either of these angles can lead to nonnegligible discrepancies in the measured acoustic travel-time shifts, which would subsequently lead to inaccurate inversions of the meridional circulation. 
In our processing, we adopt the \hmi\ $P$-angle determined from the 2012 Venus transit and two subsequent Mercury transits \citep{Hoeksema2018}. 
The $B_0$-angles for both instruments and $P$-angle for GONG are taken directly from the headers of individual Dopplergrams. 
While we cannot completely rule out the possibility of nonnegligible errors in one or more of these parameters, the data have been prepared to the best of our knowledge.

\begin{figure}
    \centering
    \includegraphics[width=0.80\linewidth]{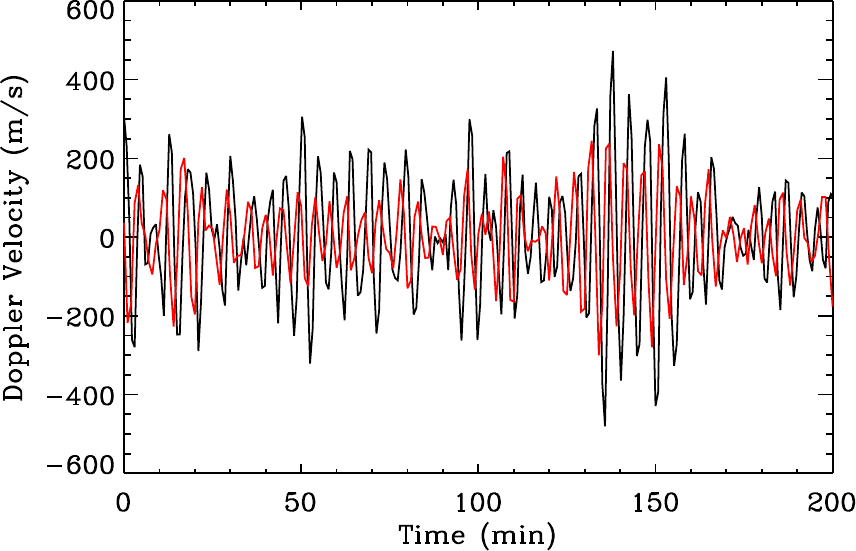}
    \caption{Simultaneous time sequences obtained by \hmi\ (dark) and GONG (red) at the same disk location that is randomly selected. }
    \label{time_sequence}
\end{figure}

\section{Relative Phase Shifts Across Solar Disk}
\label{sec3}

First, we compare the oscillatory phases in the Doppler-velocity sequences acquired by \hmi\ and GONG.

\subsection{Measuring Relative Phase Shifts}
\label{sec31}

As described in Section~\ref{sec2}, the \hmi\ and GONG datasets have been prepared in the same manner, and both cover the same time period, ideal to be used to compare the oscillatory phases of the observed Doppler velocities. 
It should be noted that the two datasets have different temporal cadences and were not acquired simultaneously. 
However, as will be shown later, our phase-shift computation does not require strict simultaneity between the two datasets, and therefore no temporal interpolation is necessary. 
Nevertheless, because we perform pixel-to-pixel cross-correlation, it is essential to ensure that the two data maps are precisely co-aligned and that the remapped pixel sizes are identical.

The procedure of calculating the relative phase shifts is as follows. 
At a given location $\mathbf{r}$, we extract one time sequence from the prepared \hmi\ data, denoted as $\psi_\mathrm{H}(\mathbf{r}, t)$, and another time sequence from the prepared GONG data, denoted as $\psi_\mathrm{G}(\mathbf{r}, t)$. 
We then compute their relative oscillatory phase shifts in the Fourier domain following the equation:
\begin{equation}
\mathcal{R}(\mathbf{r},\nu) = \widehat{\psi_\mathrm{H}(\mathbf{r}, t)} \ \widehat{\psi_\mathrm{G}(\mathbf{r}, t)^\dagger},
\end{equation}
where $\mathcal{R}(\mathbf{r},\nu)$ denotes the cross-correlation in the Fourier domain at location $\mathbf{r}$ and frequency $\nu$, $\widehat{\psi(\mathbf{r}, t)}$ indicates the Fourier transform of the time sequence $\psi(\mathbf{r}, t)$, and the symbol $^\dagger$ denotes taking the complex conjugate.

The phase shifts in the frequency range of $\nu_1$ to $\nu_2$ are then calculated as
\begin{equation}
\delta\phi_{\nu_1}^{\nu_2} (\mathbf{r}) = \arg \left( \left\langle \mathcal{R}(\mathbf{r}, \nu) \right\rangle_{\nu_1}^{\nu_2} \right),
\end{equation}
where $\arg$ denotes the argument (phase) of a complex number, and $\langle \cdot \rangle_{\nu_1}^{\nu_2}$ represents the arithmetic average over the frequency range $\nu_1 \le \nu \le \nu_2$. 
This method of computing frequency-dependent phase shifts has been frequently used in recent studies \citep[\eg,][]{Chen2018, Zhao2022, Waidele2023, Zhao2025}.
The phase shifts can be converted into time shifts following: 
\begin{equation}
    \delta\tau (\nu) = \frac{\delta\phi (\nu)} {2\pi\nu}.
\end{equation}
However, one needs to note that the $\delta\phi$ and $\delta\tau$ here represent phase shifts and time shifts from a same solar location but observed by different instruments.
They should not be confused with the travel-time shifts to be discussed in Section~\ref{sec4}, which are measured from different solar locations. 

\begin{figure}
    \centering
    \includegraphics[width=0.60\linewidth]{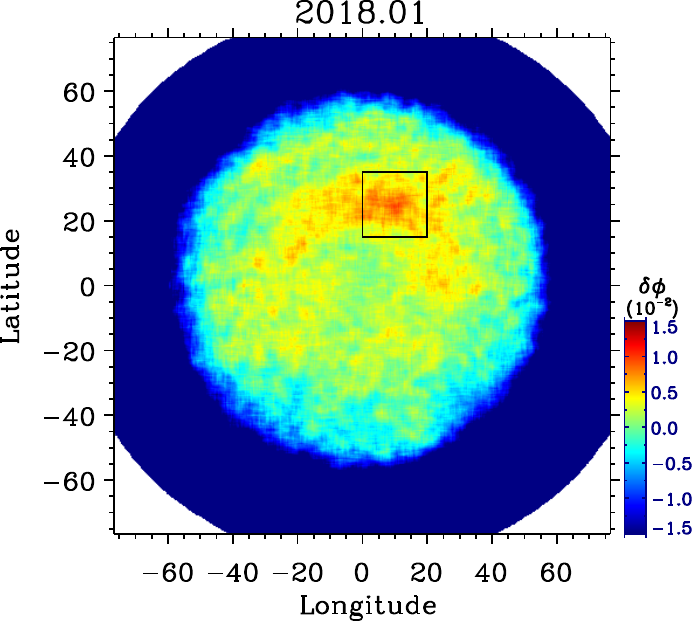}
    \caption{Map of relative phase shifts $\delta\phi(\mathbf{r})$ between \hmi\ and GONG Dopplergrams, obtained between the frequency of $3.0-4.0$\,mHz for 2018 January. 
    Note that the values are saturated near the limbs, but these areas are typically out of our interest in making meridional circulation measurements.
    The dark box indicates the location of the major phase anomalies, inside which the values are averaged for the temporal evolution shown in Figure~\ref{anomaly_evolv}. }
    \label{phase_anomaly}
\end{figure}

The same calculation procedure is applied to all other locations in the map, resulting in a two-dimensional map of phase shifts $\delta\phi (\mathbf{r})$. 
To improve the signal-to-noise ratio, the calculation is performed on the data of each day in one month, and the results from all the days are averaged for one phase-shift map for that month.
Figure~\ref{phase_anomaly} exhibits the map of $\delta\phi(\mathbf{r})$ derived for 2018 January over the frequency range of $3.0 - 4.0$\,mHz.

\begin{figure}
    \centering
    \includegraphics[width=1.0\linewidth]{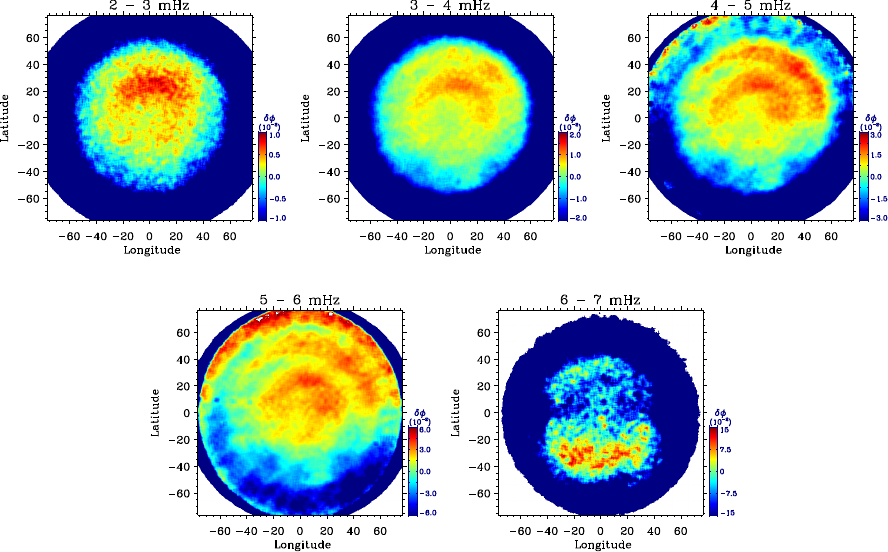}
    \caption{Map of relative phase shifts $\delta\phi(\mathbf{r})$ between \hmi\ and GONG Dopplergrams, obtained for 2018 January for the frequency range of $2.0-3.0$, $3.0-4.0$, $4.0-5.0$, $5.0-6.0$, and $6.0-7.0$\,mHz, respectively.
    Note that the values are saturated near the limbs for all panels.}
    \label{freq_dep}
\end{figure}

\subsection{Area of Phase Anomalies}
\label{sec32}

The relative phase-shift map $\delta\phi(\mathbf{r})$ measured between the \hmi\ and GONG observations contain at least three distinct components. 
The first is a uniform phase shift resulting from the difference in recording times between the two datasets, which are embedded in the Dopplergram file headers. 
We remove a uniform time offset from the computed $\delta\phi(\mathbf{r})$ maps to account for this difference. 
The second component arises because \hmi\ and GONG measure Doppler velocities using different spectral lines. 
It is expected that different lines, forming at different atmospheric heights, would introduce additional phase shifts \citep{Zhao2022, Chen2025}.
This second component thus introduces a systematic axisymmetric CtoL variation in the $\delta\phi(\mathbf{r})$ map.
Beyond these two components, the remaining signal is expected to be approximately uniform across the disk, since both instruments observe the same solar oscillations.
However, we identify a third component, consisting of localized phase-shift anomalies.
It is this component that we focus on in this work, as we believe it contributes to the discrepancies in the meridional-circulation measurements.

As shown in Figure~\ref{phase_anomaly}, which presents the $\delta\phi(\mathbf{r})$ map computed for the frequency range of 3.0 -- 4.0\,mHz, a prominent phase-shift anomaly is observed spanning approximately the longitude of $0\degr$ to $20\degr$ and latitude of $15\degr$ to $35\degr$, delimited by a box in the plot. 
The anomalies have an amplitude of approximately 7.0\,mrad (1\,mrad $=10^{-3}$\,radian), corresponding to a travel-time shift of approximately 0.3\,sec. 
These anomalies directly affect the acoustic travel-time measurements: any travel times measured for acoustic waves that either originate or end within the anomalous area are subject to artificial travel-time shifts. 

Furthermore, the phase anomalies exhibit a frequency dependence. 
Figure~\ref{freq_dep} displays the $\delta\phi(\mathbf{r})$ maps obtained for frequency bands spanning 2.0 to 7.0\,mHz with 1.0\,mHz intervals. 
Except for the high-frequency band of $6.0 - 7.0$\,mHz, all other frequency bands show phase anomalies located in approximately the same area. 
While the amplitudes of the phase shifts vary across frequency bands, the corresponding travel-time shifts remain relatively consistent. 
This implies that the phase anomalies likely originate from a discrepancy in observational timing between the two instruments in the anomalous area.
Because oscillatory signals are strongest and the measurements give the best signal-to-noise ratio in the $3-4$\,mHz band, below we will only present results obtained in this frequency band. 

\subsection{Temporal Evolution of the Phase Anomalies}
\label{sec33}

\begin{figure}
    \centering
    \includegraphics[width=1.0\linewidth]{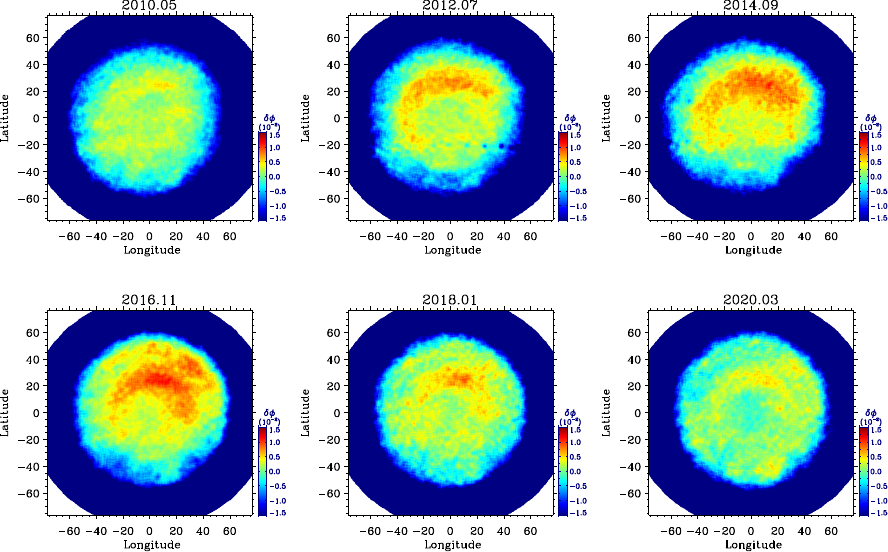}
    \caption{Map of relative phase shifts $\delta\phi(\mathbf{r})$ between \hmi\ and GONG Dopplergrams, obtained for selected months of selected years, with an intention to cover the analyzed data period and different months of the year.
    Note that the values are saturated near the limbs for all panels.}
    \label{all_years}
\end{figure}

\begin{figure}
    \centering
    \includegraphics[width=0.95\linewidth]{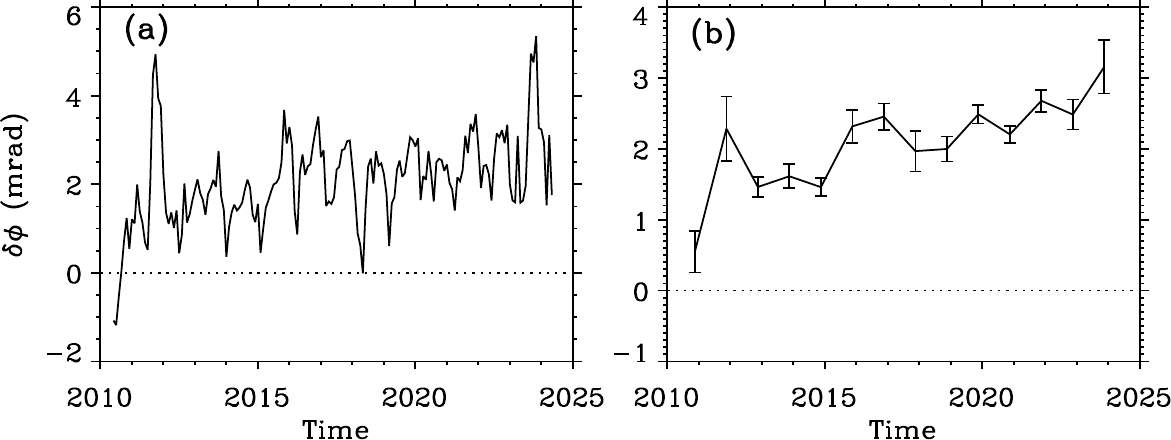}
    \caption{(a) Temporal evolution of the phase anomalies measured between the \hmi\ and GONG Dopplergrams, averaged over the area indicated by the box in Figure~\ref{phase_anomaly} for the frequency range of $3.0 - 4.0$\,mHz for each calendar month.
    (b) Annual variation of the phase anomalies, averaged from points in (a) using a 1-year window, with error bars representing the standard errors. }
    \label{anomaly_evolv}
\end{figure}

The analyzed dataset spans 14 years, during which it is reasonable to expect that the phase anomaly has undergone some variations, either due to instrumental changes or magnetic variations with the solar cycle.
Figure~\ref{all_years} presents $\delta\phi(\mathbf{r})$ maps for selected years across the full time span, as well as for selected months, in order to examine possible seasonal influences. 
Figure~\ref{anomaly_evolv}a shows the temporal evolution of the phase anomaly, which is obtained by averaging the phase shifts within frequency of $3-4$\,mHz in the anomalous area indicated by the box in Figure~\ref{phase_anomaly}, with a monthly cadence during the entire analysis period. 
Figure~\ref{anomaly_evolv}b shows the anomaly after an annual averaging.
It can be seen that the anomaly shows an increasing trend with time, but with substantial fluctuations from time to time.
The anomaly does not show apparent correlation with the solar activity cycle, but seems to vary with the time of the year.


\subsection{Phase-Shift Trend Across Disk}
\label{sec34}

As noted above, Figures~\ref{phase_anomaly} -- \ref{all_years} all carry systematic CtoL variations between the two sets of observations, primarily arising from the use of different spectral lines in the observations, among possibly other factors. 
It is interesting to examine the overall trend of the phase-shift variations between the instruments after removing an azimuthally-averaged CtoL difference, \ie, the axisymmetric component, from the phase maps.

\begin{figure}[!t]
    \centering
    \includegraphics[width=0.95\linewidth]{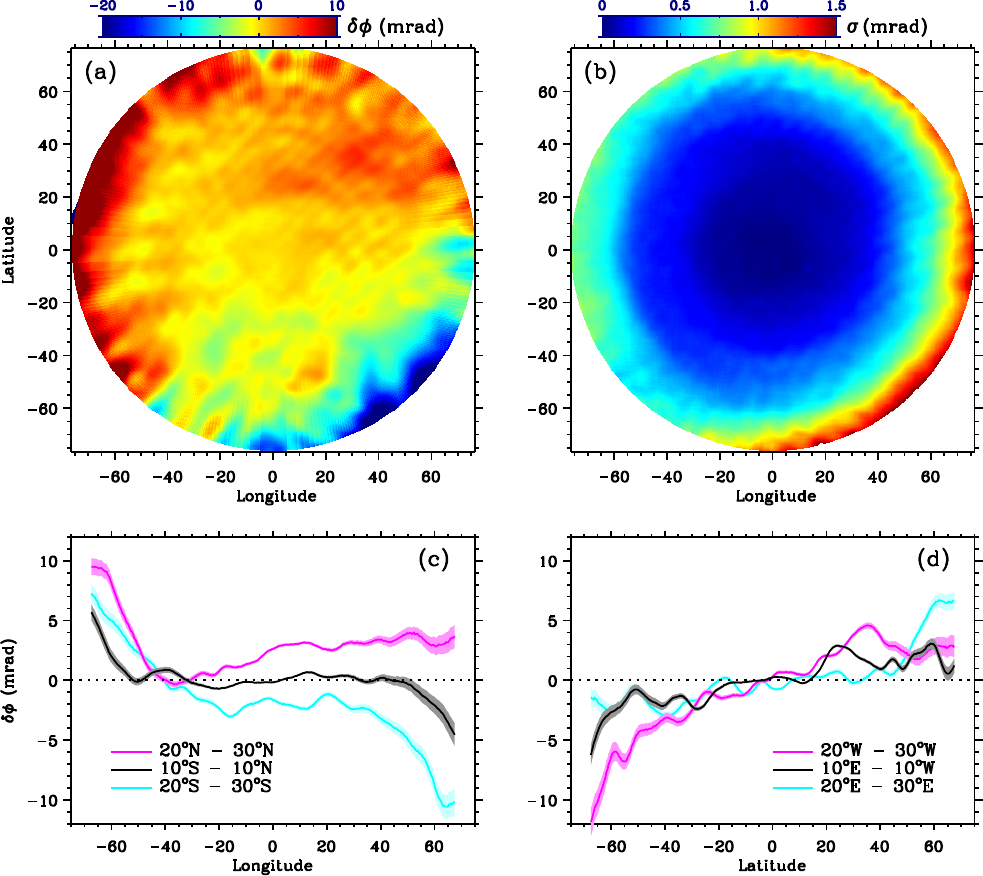}
    \caption{(a) Phase-shift map between \hmi\ and GONG Dopplergrams averaged between $3.0 - 4.0$\,mHz throughout the entire analysis period of 14 years, displayed after an azimuthally-averaged CtoL variation is removed.
    (b) Map of standard errors corresponding to phase-shift map shown in (a), estimated from monthly measurements throughout the 14-year span. 
    (c) Phase shifts averaged in the latitudinal bands of $30\degr$S -- $20\degr$S (cyan), $10\degr$S -- $10\degr$N (dark), and $20\degr$N -- $30\degr$N (magenta), displayed as functions of longitude.
    Shades of different colors indicate the range of errors.
    (d) Phase shifts averaged in the longitudinal band of $30\degr$E -- $20\degr$E (cyan), $10\degr$E -- $10\degr$W (dark), and $20\degr$W -- $30\degr$W (magenta), displayed as functions of latitude. 
    Note that the map in (a) is in Postel-projected coordinate and the averaging bands to obtain curves in panels (c) and (d) are thus not in heliographic coordinate.}
    \label{ave_anomalies}
\end{figure}

In Figure~\ref{ave_anomalies}, panel (a) shows the 14-year-averaged $\delta\phi(\mathbf{r})$ map after the removal of the CtoL variation, and panel (b) shows the standard errors estimated from monthly phase-shift maps spanning the entire analysis period.
Panel (c) illustrates the variation trend with longitude for selected latitudinal bands, and panel (d) presents the trend with latitude in selected longitudinal bands. 
Despite an expectation of a nearly uniform phase shift across the disk, a noticeable and alarming trend emerges: the phase shift gradually decreases from the northeastern quadrant to the southwestern quadrant. 
The phase-anomaly patch discussed in the above sections remains prominent after the long-term averaging.
In addition, diagonal stripe patterns oriented at approximately $45\degr$ to the horizontal are evident, mostly in the northern hemisphere. 
These features are likely owing to the interference patterns present in the \hmi\ filtergrams \citep{Couvidat2012}, which propagate into the Dopplergrams and subsequently into the oscillatory phase measurements. 
It is interesting to note that the interference patterns appear nearly perpendicular to the gradient of the phase-shift maps, although it is unclear whether these are related.

The gradual across-disk phase-variation trend can be interpreted in helioseismic measurements as a southward flow, stronger on the western side of the central meridian than on the eastern side (the $\delta\phi(\mathbf{r})$ shows larger gradient on the western side than on the eastern side, as indicated in Figure~\ref{ave_anomalies}d). 
It may also manifest as apparent westward flows in the southern hemisphere and smaller-magnitude eastward flows in the northern hemisphere (the $\delta\phi(\mathbf{r})$ shows clear gradient in the southern hemisphere but little gradient in the northern hemisphere near the central meridian, as indicated in Figure~\ref{ave_anomalies}c), potentially producing a long-term bias of that the southern hemisphere appears to rotate faster than the northern hemisphere.
For a comparison, the phase-shift trend shows little variation along the equatorial area within $\pm50\degr$ from the central meridian (gray curve in Figure~\ref{ave_anomalies}c), but shows clear gradient along the central meridian even after the CtoL effect is removed (gray curve in Figure~\ref{ave_anomalies}d). 

Here, it is important to emphasize two points. 
First, the phase maps in Figure~\ref{ave_anomalies} represent {\it relative} phases between the two instruments. 
Thus, they do not directly indicate which, if either, instrument is responsible for the discussed phase anomalies, although it is highly plausible that systematic errors exist in both datasets, more or less. 
Second, the phase shifts are measured within specific frequency bands without decomposition into individual wavenumbers, making it difficult to assess the depths at which the inferred rotational or meridional flows may be most affected. 
It is also possible that the unreliability is confined to a narrow range of wavenumbers, thus a limited depth range, without significantly impacting the helioseismic inversion results throughout the broader convection zone.

\section{Travel-Time Measurements}
\label{sec4}
In Section~\ref{sec3}, we measured the relative phase shifts between the oscillatory signals observed by \hmi\ and GONG, in which an area with strong phase-shift anomalies and an across-disk phase-variation trend are identified.
However, how those phase anomalies and the gradual phase trend are related to our meridional-circulation measurements is not immediately clear.
In this section, we investigate how travel-time measurements from the two instruments are different from each other and how they are connected with the relative phase shifts reported in Section~\ref{sec3}.

\subsection{Offset in Travel-time Shifts}
\label{sec41}

\begin{figure}
    \centering
    \includegraphics[width=0.80\linewidth]{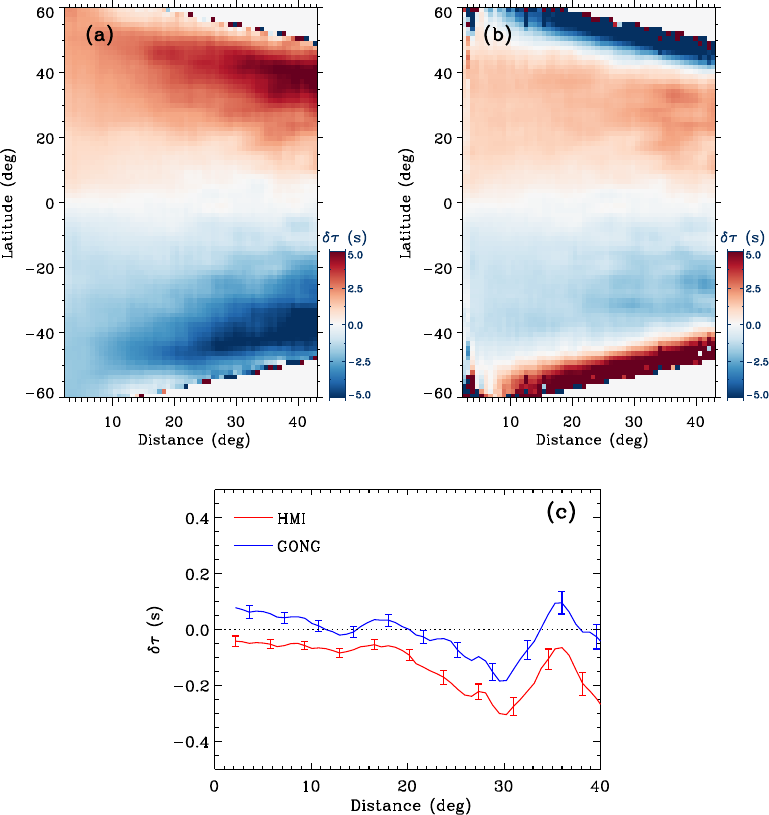}
    \caption{(a) Acoustic travel-time shifts, $\delta\tau(\Delta, \lambda)$, measured along the Sun's North-South direction using \hmi\ data for the period of 2010.05.01 -- 2024.04.30, and displayed as a function of measurement distance $\Delta$ (corresponding to measurement depth) and latitude $\lambda$.
    The travel-time shifts include both the helioseismic CtoL effect and the meridional-circulation-induced shifts, without the former subtracted.
    (b) Same as (a) but from GONG observations during the same period.
    (c) Comparison of $\delta\tau$ measured from \hmi\ and GONG data and averaged from the latitudinal band of $20\degr$S -- $20\degr$N.}  
    \label{full_disk_dt}
\end{figure}

Consistent with the methods in another study \citep{Chen2025}, travel-time shifts along the Sun’s north-south direction are measured using both \hmi\ and GONG Dopplergrams within the central-meridian region. 
The travel-time shifts, $\delta\tau(\Delta, \lambda)$, are defined as the difference between the travel times of northward- and southward-propagating waves, measured between a pair of locations situated along a same longitudinal line but separated by a distance $\Delta$, with the midpoint located at latitude $\lambda$. 
The measurements are performed within a longitudinal band of $30\degr$ wide, spanning $\pm15\degr$ on either side of the central meridian.
Figures~\ref{full_disk_dt}a and \ref{full_disk_dt}b present the travel-time shifts, $\delta\tau (\Delta, \lambda)$, obtained from 14 years of \hmi\ and GONG data, respectively, displayed as functions of measurement distance $\Delta$ and latitude $\lambda$. 
It should be noted that the dominant signals in these two panels arise from systematic CtoL effects.
Because the two instruments observe different spectral lines and have different spatial resolutions, they probe different line-formation heights and experience different degrees of geometric foreshortening; consequently, their CtoL effects differ substantially. 
This difference is particularly pronounced at higher latitudes, where the $\delta\tau$ measured from GONG data even reverses sign. 
However, as shown by \cite{Chen2025}, once the CtoL contribution is removed, the meridional-flow-induced $\delta\tau$ remains consistent between the two datasets. 

Figure~\ref{full_disk_dt}c compares the \dtdl\ values averaged over the latitudinal band between $20\degr$S to $20\degr$N.
Because the CtoL effect is presumably anti-symmetric about the equator, this averaging procedure effectively cancels the effect, allowing a direct comparison between the \hmi\ and GONG results without the need for explicit correction. 
A clear offset is seen between two sets of measurements, and the offset is significantly larger than typical error bars.
This is consistent with earlier reports \citep{Gizon2020}.
It should also be noted that this offset does not depend significantly on the selection of widths of latitudinal band. 

\begin{figure}
    \centering
    \includegraphics[width=0.80\linewidth]{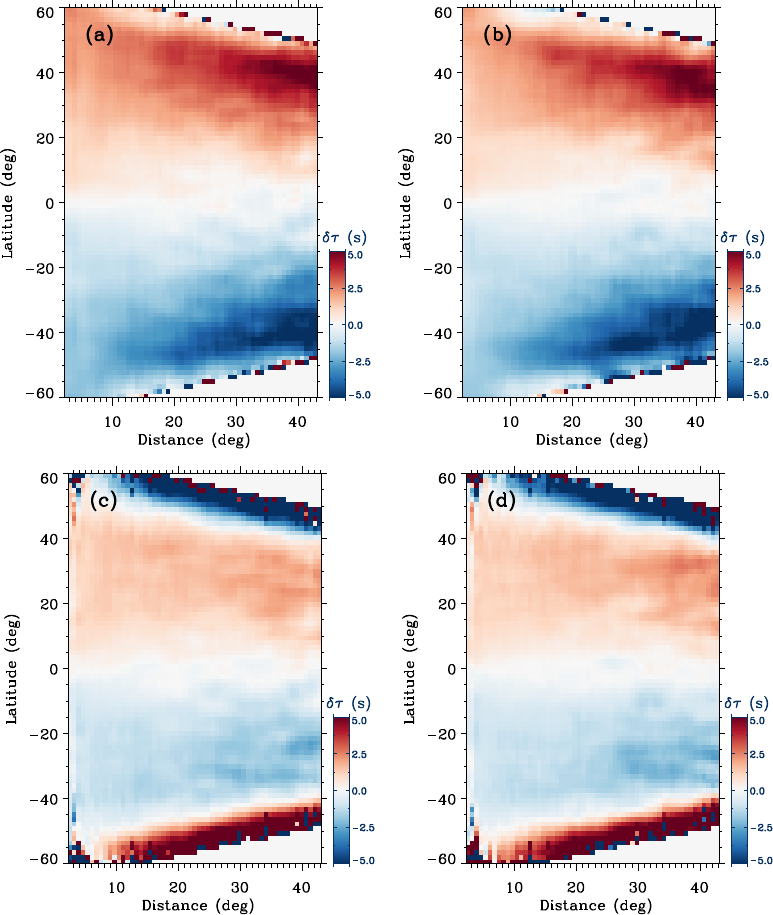}
    \caption{(a) Travel-time shifts, \dtdl, measured using \hmi\ data on the eastern side of the central meridian. 
    (b) The \dtdl\ measured using \hmi\ data on the western side the central meridian. 
    (c) The \dtdl\ measured using GONG data on the eastern side of the central meridian. 
    (d) The \dtdl\ measured using GONG data on the western side of the central meridian.}
    \label{east_west}
\end{figure}

\begin{figure}
    \centering
    \includegraphics[width=1.0\linewidth]{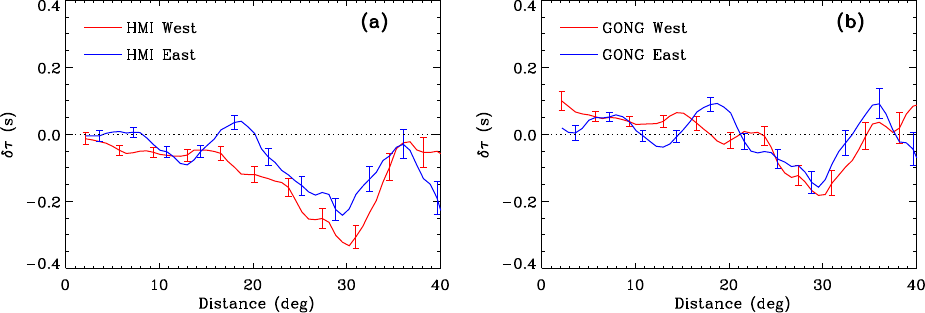}
    \caption{(a) Comparison of travel-time shifts \dt\ measured using \hmi\ data on the eastern and western sides of the central meridian, after the shifts are averaged over the latitudinal band of $20\degr$S -- $20\degr$N. 
    (b) Same as panel (a), but obtained from using GONG data.}
    \label{dt_instr_oriented}
\end{figure}

\begin{figure}
    \centering
    \includegraphics[width=1.0\linewidth]{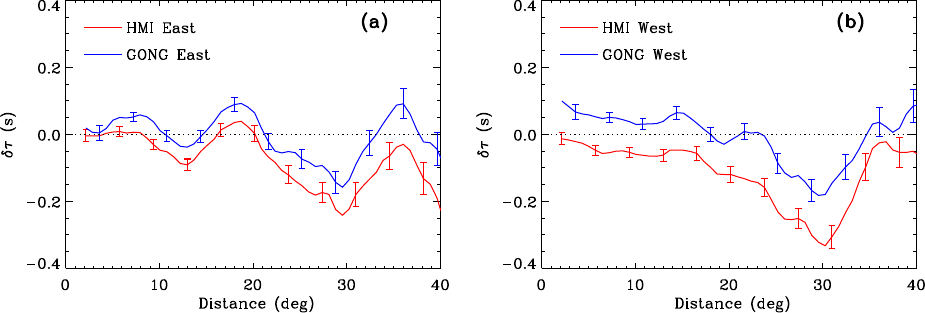}
    \caption{(a) Comparison of travel-time shifts \dt\ measured using \hmi\ and GONG data on the eastern side of the central meridian, after the shifts are averaged over the latitudinal band of $20\degr$S -- $20\degr$N. 
    (b) Same as panel (a) but for the measurements on the west side of the central meridian. }
    \label{dt_ew_oriented}
\end{figure}

\subsection{Discrepancies in Eastern and Western Sides of Central Meridian}
\label{sec42}

We then perform separate measurements of \dtdl\ on the eastern and western sides of the central meridian, each covering a longitudinal span of $15\degr$. 
This procedure results in four distinct datasets: \dthw\ and \dthe, measured using \hmi\ data on the western and eastern sides of the central meridian; \dtgw\ and \dtge, measured using GONG data on the western and eastern sides of the central meridian. 
These measurements from 14 years of observations are presented in Figure~\ref{east_west}.
In principle, one would not expect \dthw\ to differ significantly from \dthe, nor \dtgw\ from \dtge, particularly after 14 years of averaging, since measurements from both sides of the central meridian should adequately represent the Sun’s meridional circulation. 
However, our results indicate the otherwise.

Figure~\ref{dt_instr_oriented} shows the comparisons between \dthw\ and \dthe, and between \dtgw\ and \dtge, after averaging the measurements over the $20\degr$S to $20\degr$N latitudinal band. 
It is evident that \dthw\ differs significantly from \dthe, with the differences exceeding the estimated error bars in most places even after averaging over 14 years. 
The discrepancy between \dtgw\ and \dtge\ appears smaller than that between \dthw\ and \dthe, but is statistically significant in many places, too.

Figure~\ref{dt_ew_oriented} further compares the averages of \dt\ from the \hmi\ and GONG datasets for the western and eastern sides, respectively. 
Both comparisons reveal substantial discrepancies between measurements on the two sides, with the difference being smaller on the eastern side than on the western side though.

The measurements presented in Figures~\ref{east_west} -- \ref{dt_ew_oriented} lead to at least two points. 
First, the results obtained from both instruments exhibit inconsistencies on either side of the central meridian, indicating that neither observation is entirely free of systematic errors, although the discrepancies are more pronounced in the \hmi\ observations. 
Second, when comparing the measurements on the same side of the central meridian using observations from two different instruments, substantial differences can be seen, although a better agreement is found on the eastern side than on the western side.

\subsection{Temporal Evolution of the Discrepancies}
\label{sec43}

As discussed above, the measured travel-time differences \dtdl\ on the western and eastern sides of the central meridian differ for both \hmi\ and GONG data, and the \dtdl\ obtained from the same side of the meridian but from different instruments also show substantial discrepancies. 
It is useful to examine how these discrepancies evolve over the 14-yr analysis period.

As seen in Figures~\ref{dt_instr_oriented} and \ref{dt_ew_oriented}, the discrepancies between the eastern- and western-side measurements for a same instrument, or same-side measurements from different instruments, change with the measurement distance $\Delta$. 
Here, we use $\langle\delta\tau^\mathrm{W} - \delta\tau^\mathrm{E} \rangle$ averaged over $\Delta = 5\degr - 10\degr$ to evaluate the temporal evolution of the eastern-western asymmetry, and use $\langle\delta\tau^\mathrm{H} - \delta\tau^\mathrm{G} \rangle$, also averaged over $\Delta = 5\degr - 10\degr$, to evaluate the temporal evolution of the instrumental discrepancies on either side of the central meridian.

\begin{figure}
    \centering
    \includegraphics[width=0.99\linewidth]{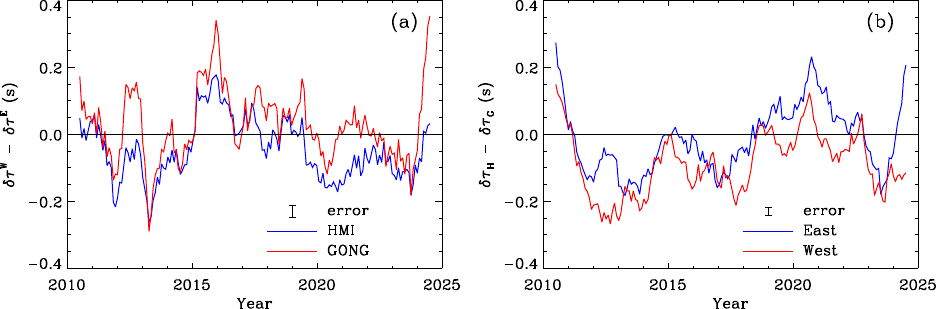}
    \caption{(a) Temporal evolution of the eastern-western travel-time asymmetry measured from \hmi\ and GONG Dopplergrams separately. 
    (b) Temporal evolution of the travel-time differences between \hmi\ and GONG measurements, displayed separately for the eastern and western side of the central meridian. 
    Sizes of typical measurement errors are illustrated in the legend area of each panel. }
    \label{diff_WE_eq}
\end{figure}

Figure~\ref{diff_WE_eq}a presents the temporal evolution of the eastern-western asymmetry measured separately from \hmi\ and GONG. 
Overall, the \hmi\ results exhibit a variation trend similar to that from GONG, although consistently smaller in magnitude. 
The clear correlation between the two sets of measurements suggests that the eastern-western asymmetry cannot be attributed solely to instrumental effects; otherwise, one would not expect the two independent measurements to correlate so well. 
On the other hand, it is also unlikely that the asymmetry reflects genuine solar variability, since each data point represents a one-year running average (except at the beginning and end of the curves). 
Given the Sun’s rapid rotation, the eastern and western sides of the central meridian should not differ significantly after such temporal averaging. 
This may imply the possible presence of unknown factors that influence helioseismic measurements -- factors that may be coupled with the true signals but are not of the same origin.

Figure~\ref{diff_WE_eq}b shows the temporal evolution of the differences between \hmi\ and GONG measurements for both the eastern and western sides of the central meridian. 
As expected, the curves are predominantly negative, reflecting the systematic offset described in Section~\ref{sec41}. 
It is also expected that the western-side measurements are more negative than the eastern-side ones, consistent with the stronger phase-shift trend observed on the western side (Figure~\ref{ave_anomalies}). 
The close correlation between the two curves is understandable, as both correspond to measurements of the same Sun, albeit obtained at slightly different times and subject to minor solar evolution between the two hemispheric observations.

It should be evident that most \dt\ measurements presented in this Section is consistent with the phase-shift $\delta\phi$ measurements presented in Section~\ref{sec3}. 
The eastern and western side \dt\ asymmetry likely arises from the gradual phase-shift variations across the disk and is also due to the apparent phase-anomaly area in the northwestern quadrant of the solar disk.
The travel-time offset seen in Figure~\ref{full_disk_dt}c is likely caused by the gradual phase variations across the disk seen in Figure~\ref{ave_anomalies}a.
The temporal evolution the \hmi\ and GONG discrepancies (Figure~\ref{diff_WE_eq}b) is consistent with the temporal evolution of the phase anomaly shown in Figure~\ref{anomaly_evolv}.

\section{Discussion}
\label{sec5}
To reconcile the discrepancies in meridional-circulation measurements derived from \hmi\ and GONG observations, it is crucial to identify the underlying sources of these differences. 
An important step toward this goal is a direct comparison of high-cadence Dopplergrams obtained from both instruments.
In this study, we measure the relative oscillatory phase shifts between the two datasets at the same solar locations and investigate their temporal evolution. 
We also examine the travel-time shifts along the north-south direction on both sides of the central meridian.

Our analysis reveals an area in the northwestern quadrant of the solar disk that exhibits pronounced phase anomalies  (Figure~\ref{phase_anomaly}).
These anomalies persist throughout the entire analysis period, displaying a general increase over time with substantial temporal fluctuations (Figure~\ref{anomaly_evolv}). 
After removing the axisymmetric component, the long-term averaged phase shifts show a clear, gradual decrease from the northeastern to the southwestern quadrant (Figure~\ref{ave_anomalies}). 
This systematic phase gradient would introduce an apparent southward flow that is stronger on the western side of the central meridian than the eastern, and would also lead to an apparent faster rotation in the southern hemisphere than in the northern.
The travel-time offset between the \hmi\ and GONG measurements seen in Figure~\ref{full_disk_dt}c is likely due to the same factor.

All these findings raise concerns for global-scale studies of the Sun’s internal rotation and meridional circulation. 
However, it is important to keep in mind that both the phase anomalies and the gradual phase-shift patterns are {\em relative} measurements between the \hmi\ and GONG observations. 
Consequently, it remains difficult to determine which instrument contributes what effects to which regions on the solar disk. 
In addition, this phase-shift analysis is performed in the frequency domain without accounting for wavenumber information; therefore, the depths to which these phase anomalies may influence helioseismic inferences cannot be readily determined.
However, the time-distance measurements on either side of the central meridian (Figure~\ref{dt_instr_oriented}) seem to show that the discrepancies caused by the phase anomalies are rather uniform across most of the measurement distances, indicating that the effect from the phase anomalies is wide across most depths in the convection zone.

Our analysis also reveals that, for both instruments, the travel-time shifts measured on the eastern and western sides of the central meridian are not fully consistent with each other (Figure~\ref{dt_instr_oriented}), despite the expectation of close agreement.
Discrepancies are also evident between the \hmi\ and GONG measurements on the same sides of the central meridian  (Figure~\ref{dt_ew_oriented}), although they are less pronounced on the eastern side. 
These results suggest that both instruments introduce systematic biases, rather than random errors, which manifest as an apparent southward flow, particularly on the western side of the central meridian. 
This finding seems to imply that restricting meridional circulation measurements to the eastern hemisphere may yield improved consistency between the two sets of observations and may better represent the actual solar flows. 
However, such an approach is not much feasible, as correcting the CtoL effect remains essential. 
The CtoL correction is typically derived from travel-time measurements along the equator, which inevitably include data from both hemispheres. 
Consequently, this correction may introduce additional artifacts that further complicate the interpretation of meridional flow measurements.

The temporal evolution of the eastern- and western-side travel-time asymmetry shows a strong correlation between the \hmi\ and GONG measurements (Figure~\ref{diff_WE_eq}). 
This suggests that the observed asymmetry is not entirely instrumental in origin, but may arise from unknown factors that are coupled with genuine solar signals. 
Such an eastern-western asymmetry cannot be of solar origin, as a rapidly rotating Sun is not expected to produce a pronounced or time-varying asymmetry in these measurements.

These findings further complicate the reliable inference of meridional circulation. 
At present, most helioseismic measurements rely on using east-west travel-time measurements near the equator as a proxy for the CtoL effect, which is then subtracted from north-south measurements \cite[e.g.,][]{Zhao2013, Gizon2020}. 
To improve the signal-to-noise ratio of this proxy, it is often assumed that the CtoL effect is stationary for a long time.
However, Figure~\ref{ave_anomalies} shows that the phase anomalies exhibit an across-disk trend, and the equatorial region and the central meridian do not necessarily share the same behavior, undermining the validity of using one as a proxy for the other. 
Furthermore, the temporal evolution of the phase anomalies (Figures~\ref{anomaly_evolv} and \ref{diff_WE_eq}) suggests that the CtoL effect is not stationary over long timescales, either. 


\section{Conclusion}
\label{sec6}

In summary, our study identifies several concerning factors that may contribute to the long-standing challenges in determining the Sun’s internal meridional circulation. 
An area of strong phase anomalies and a systematic, across-disk gradually-changing trend in phase-shift variations are found in the relative phases derived from \hmi\ and GONG Dopplergrams. 
For both instruments, the north-south travel-time measurements obtained on either side of the solar central meridian show clear inconsistencies, indicating that neither dataset is free from systematic errors. 
The strong correlation between the eastern-western travel-time asymmetries measured by the two instruments further suggests that the discrepancies arise from more complex causes than simple instrumental imperfections. 
Moreover, both the phase anomalies and the eastern-western asymmetry evolve over time, posing a major challenge to their full characterization and correction, and thus to achieving an accurate determination of the meridional circulation.


\begin{acks}
\sdo\ is a NASA mission, and \hmi\ is a NASA project contracted to Stanford University.
This work utilizes GONG data obtained by the NSO Integrated Synoptic Program, managed by the National Solar Observatory, which is operated by the Association of Universities for Research in Astronomy (AURA), Inc. under a cooperative agreement with the National Science Foundation and with contribution from the National Oceanic and Atmospheric Administration. 
\end{acks}

\begin{fundinginformation}
This work was partly sponsored by NASA Heliophysics Guest Investigator program under contract number 80NSSC25K7671. J.~Z. and R.~C. were also supported by the NASA DRIVE Science Center COFFIES grant 80NSSC22M0162. 
\end{fundinginformation}

\begin{dataavailability}
The \hmi\ Dopplergrams are publicly available at \url{http://jsoc.stanford.edu}, and the GONG Dopplergrams are also publicly available at \url{https://gong.nso.edu}.
\end{dataavailability}

\begin{conflict}
The authors declare that they have no conflicts of interest.
\end{conflict}

\bibliographystyle{spr-mp-sola}
\bibliography{main}

\end{document}